\documentclass[a4paper]{article}

\usepackage{INTERSPEECH2020}

\usepackage{booktabs} 
\usepackage{multirow}
\usepackage{algorithm}
\usepackage{algorithmic}
\usepackage[font=footnotesize]{caption}

\title{Accurate Detection of Wake Word Start and End Using a CNN}
\name{Christin Jose\textsuperscript{*}\thanks{\textsuperscript{*}Equal contribution}, Yuriy Mishchenko\textsuperscript{*}\footnotemark[1], Thibaud Senechal, Anish Shah, Alex Escott, Shiv Vitaladevuni}
\address{
  Amazon Alexa Research}
\email{\{chrjse,yuriym,thibauds,anishsh,escottal,shivnaga\}@amazon.com}

\begin{document}









\maketitle
\begin{abstract}
Small footprint embedded devices require keyword spotters (KWS) with small model size and detection latency for enabling voice assistants. Such a keyword is often referred to as \textit{wake word} as it is used to wake up voice assistant enabled devices. Together with wake word detection, accurate estimation of wake word endpoints (start and end) is an important task of KWS.  In this paper, we propose two new methods for detecting the endpoints of wake words in neural KWS that use single-stage word-level neural networks. Our results show that the new techniques give superior accuracy for detecting wake words' endpoints of up to 50 msec standard error versus human annotations, 
on par 
with the conventional Acoustic Model plus HMM forced alignment. To our knowledge, this is the first study of wake word endpoints detection methods for single-stage neural KWS.
\end{abstract}

\maketitle


\section{Introduction}
Keyword spotting is the task of detecting keywords of interest in a continuous audio stream. It has been an active research area in speech recognition and applications recently. With the recent increase in the popularity of voice assistants such as Alexa, Hey Google, and Siri, KWS have attracted much attention in the context of on-device wake word (WW) spotting. Accurate detection of WW endpoints in an audio stream is an important feature of KWS in WW applications. Voice assistant enabled devices only start streaming audio to the cloud when the Keyword Spotter detects a WW, and streaming must start from the WW start point.

Conventional WW detection methods use 2-stage models comprising a first stage Deep Neural Network acoustic model (AM DNN) and a Hidden Markov Model (HMM) \cite{Panchapagesan2016MultiTaskLA, Compressedinproceedings, Sun2017AnES, Kumatani2017DirectMO, Bottleneckinproceedings}. Such a keyword spotter may also have additional classifiers after the HMM, such as an SVM, to increase the accuracy of WW detection \cite{minhua1}. These KWS naturally provide the endpoints of the WW via the HMM output. Specifically, during runtime, these systems perform Viterbi decoding of the WW senone sequences that produces the times of the WW's start and end senones in the input audio. However, this procedure can be computationally expensive, depending on the HMM topology, and training of such a keyword spotter is very complex.


A recent work investigated KWS based on a single-stage feed-forward DNN \cite{Smallfootprint1}. That DNN is trained to predict sub-keyword targets and has been shown to outperform a Keyword/Filler HMM approach. Such DNNs are also attractive for running on hardware-limited devices since the size of the model can be easily controlled to fit the devices' CPU and memory budget by changing the number of parameters in the DNN. Convolutional Neural Networks (CNNs) have also become popular for acoustic modeling and have shown improvements over the fully connected feed-forward DNNs as KWS \cite{sainath15}.

An important functionality of WW KWS is the ability to determine the endpoints of a WW in the audio. WW endpoints are used to decide which audio will be sent to the cloud, which helps to protect user privacy and reduces the cloud-side processing costs. The recent neural keyword spotting methods discussed above, aim at improving WW detection accuracy. However, the detection of WW endpoints becomes a greater problem with those methods. E.g., compared with the conventional AM+HMM approach, no such output as the HMM senone states is available in neural KWS. The Deep keyword spotter \cite{Smallfootprint1} can estimate the endpoints of the keyword in an audio based on the rise of the model's posterior corresponding to the WW sub-word labeling. However, this may not detect the endpoints of the WW very accurately.

In this paper, we consider WW spotter models designed as single-stage feed-forward CNN operating on an audio context up to a second, or word-level KWS. We introduce two methods for detecting keyword endpoints in such word-level KWS. The first method uses a second regression model trained on intermediate representations of keyword spotting CNN, in order to predict the keyword endpoints inside the input window. The second method uses a novel approach of a multi-aligned CNN model trained to detect keyword in different alignments inside the input window, such as towards the start or the end of the input window. To our knowledge, these are the first methods in the literature for keyword endpointing in word-level KWS. Likewise, the approach of multi-aligned keyword modeling is novel and may be of interest to other applications. The described methods improve standard error for keyword endpoints up to 60\% compared to a constant offset algorithm. Our methods have an equivalent standard error when compared with gold standard method of AM+HMM model. These methods will allow significantly simpler model training and inference as well as better keyword spotting accuracy.

\begin{figure*}[!htbp]
\centering
\includegraphics[width=1\textwidth]{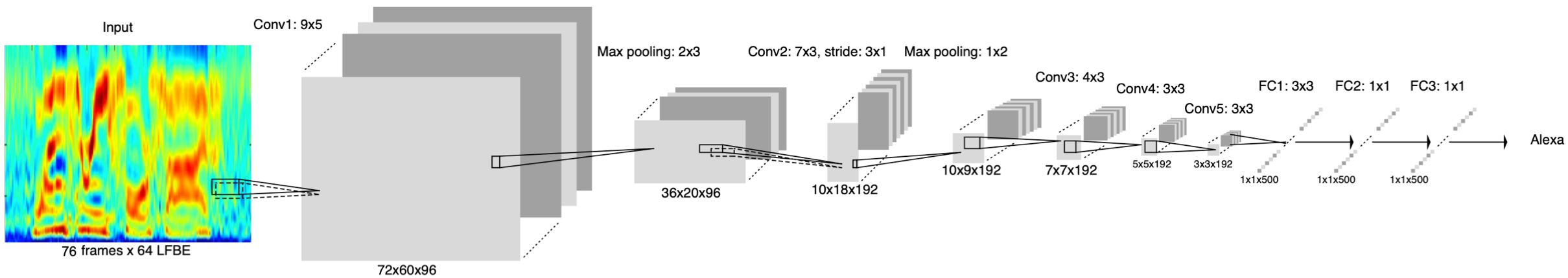}
\caption{Word level CNN architecture.}
\vspace{-5mm}
\label{fig:cnn_thibaud_diagram}
\end{figure*}

\section{Word-level Keyword Spotting model}\label{sec:wordLevel}
The word-level (WL) keyword spotter considered here is a CNN WW detector similar to the Deep KWS \cite{Smallfootprint1} and the CNN KWS \cite{sainath15}. However, differently from those, here the input window encompasses the entire audio context of one WW---the input window can range up to 1 sec. 
Specifically, the model is trained by using a set of positive examples (i.e. the audio fragments containing the WW) and negative examples (i.e. the audio fragments not containing the WW). In the WW-positive examples, the WW is consistently aligned in the long CNN input window, such as centered. The model learns to produce output representing the posterior probability of finding the WW in a given input audio fragment.

Our CNN operates on 64-dimensional Log mel Filter-Bank Energy (LFBE) features calculated over the standard 25 msec frames with a 10 msec shift. The CNN architecture is five convolutional layers plus three fully connected (FC) layers with max-pooling after the first layer and 3-stride convolution in the second layer. See Fig. \ref{fig:cnn_thibaud_diagram} for specific architecture details. 
Dropout and batch normalization are used with all hidden layers. 
The output layer is softmax over two outputs comprising ``WW" and ``non-WW". 
The model is trained using cross-entropy loss on data prepared as a mixture of WW-positive and WW-negative word-level examples, as described above, with labels in one-hot encoding. The described CNN architecture achieves superior baseline accuracy for neural WW spotting.

\section{Baseline Methods for WW endpoints detection} \label{sec:existingMethods}

Choosing baselines for the WW endpoints detection algorithm in the context of WL KWS is difficult as no methods for WW endpointing exist for that setting. Unlike ``frame-level" models such as AM+HMM (i.e. KWS that work on sub-word input), neural WL KWS by design does not offer natural means for detecting the WW endpoints in the audio, because such models are trained to detect varying-length WW in the input window uniformly. That is, one cannot know from a detection event alone where in the input window, the WW starts and ends, without additional information about at least the WW length. We consider two baselines in this work---the AM+HMM KWS as the industry's gold standard for keyword endpointing, and a constant offset method, which is the most straightforward algorithm for WW endpointing in WL KWS. 

\subsection{AM+HMM endpointing}
The 2-stage AM+HMM KWS \cite{Panchapagesan2016MultiTaskLA, Compressedinproceedings, Sun2017AnES, Kumatani2017DirectMO, Bottleneckinproceedings} is de-facto the golden standard for keyword endpointing in the industry. In that algorithm, the posteriors are produced by an Acoustic Model DNN for a set of senones, based on an input audio stream, and an HMM is tuned to force-align a sequence of senones expected in the keyword to those detected. This is the classical approach used in ASR. The keyword endpoints are naturally produced in that algorithm as the times of the first and the last senones in the HMM state sequence corresponding to a keyword detection \cite{HMMpatent}. In this paper, we use a 2-stage WW model from \cite{minhua1} for such endpointing baseline.

\subsection{Constant offset endpointing}
Constant offset method is the simplest algorithm for detecting WW endpoints in WL KWS.
In this approach, we depart from the view that the WW has a relatively small variation in length, such as due to pronunciation by different speakers or speed of speech differences. For example, for ``Alexa" the 
10-90\textsuperscript{th} percentile 
variability in keyword length across speakers is 500 to 900 msec. In that setting, we may estimate the WW start and end points by using a suitably chosen constant offset from the time of the WW detection event in WL KWS, given known typical WW duration and expected alignment in the keyword spotter input window. 
For different WWs, an optimal offset can be chosen based on a measured mean or median duration of those WWs. 
While simple in principle, the accuracy of this method may not be satisfactory for WWs whose length can vary greatly from the mean or median value.

\section{WW endpoints detection in WL KWS}

\subsection{WW start-end regression model method}
\begin{figure}[!htb]
\centering
\vspace{-3mm}
\includegraphics[width=0.47\textwidth]{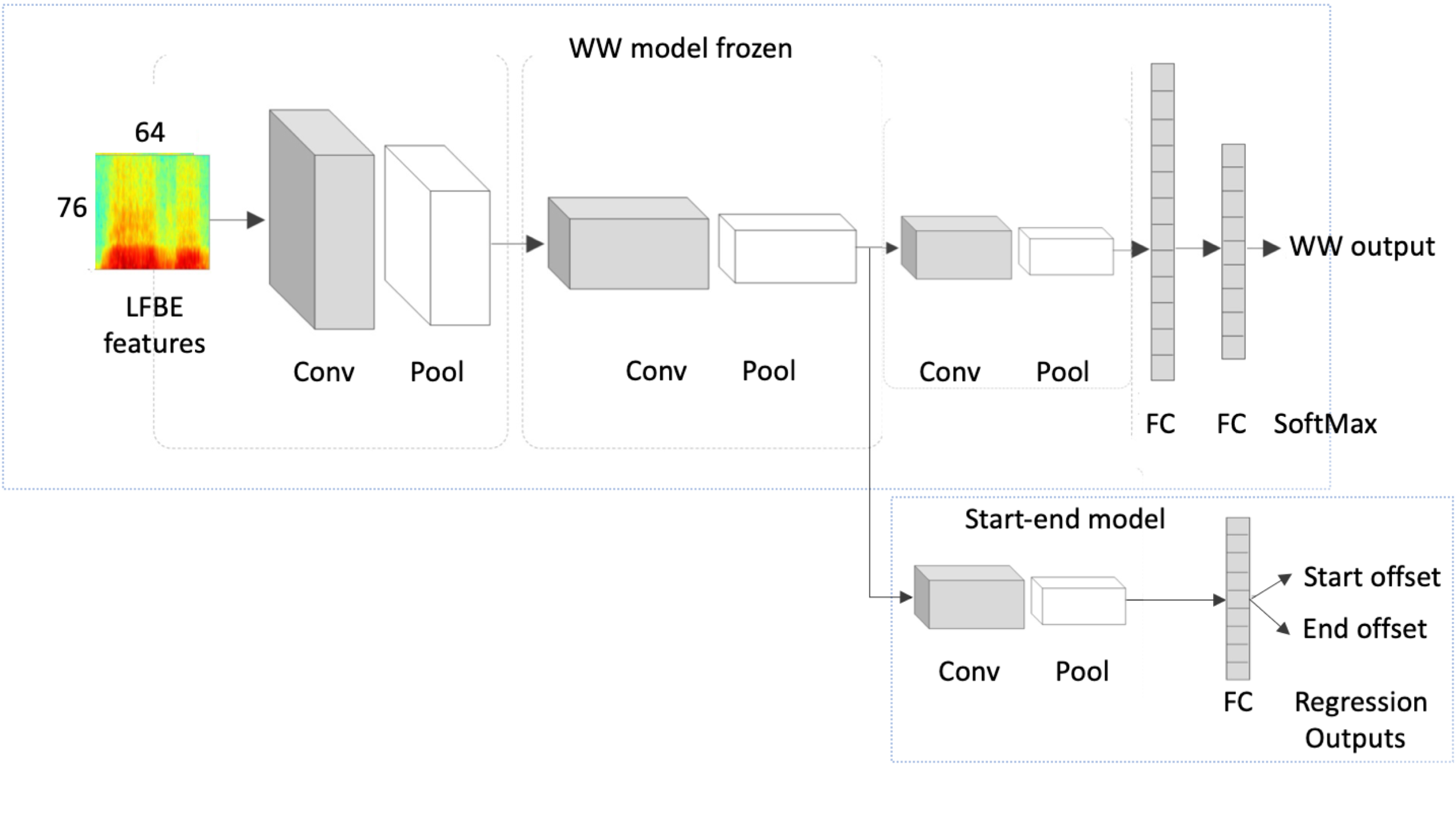}
\vspace{-8mm}
\caption{Endpointing in WL KWS using second start-end regression model.}
\label{fig:cnn-start-end}
\vspace{-2mm}
\end{figure}

In this method, we add a second regression model that runs in parallel with the main detector WL CNN. The regression model uses intermediate feature representations from the hidden layers of the WL CNN as its input and is trained to output the relative offsets of the WW endpoints in the input window against ground truth, Fig \ref{fig:cnn-start-end}.
The ground-truth is prepared via pseudo-labels produced by the AM+HMM KWS \cite{minhua1}. 
The model is trained using the mean square error loss.

More specifically, we first train the main WW detector CNN using data prepared as a mixture of WW-positive and WW-negative WL examples, as described in Section \ref{sec:wordLevel}. 
We then add the second branch of the start-end regression model. The key idea is that the intermediate representations from the WL CNN can allow predicting the positions of WW endpoints.  We experimented with training the main detector CNN and the start-end regression model simultaneously, in a multi-task manner, or freezing the main CNN weights and training the start-end model separately. We found that freezing the main CNN was the best for WW detection accuracy.

The start-end regression model comprises one convolution layer of dimensions (5,5,200) and one fully connected layer with two outputs. 
We experimented with the detector CNN's hidden layer serving as the input for the start-end regression model and found the convolution layer 4 to produce the best results. 
The two outputs encode the start and end offsets of WW inside the input window measured in units relative to the input window length. That is, output 0 corresponds to the beginning of the input window, and output 1 corresponds to the end of the input window. The WW can be longer than the input window. In that case, the start can be negative, and the end can be greater than 1.

\subsection{Multi-aligned output WW model}
\begin{figure}[!htb]
\centering
\includegraphics[width=0.48\textwidth]{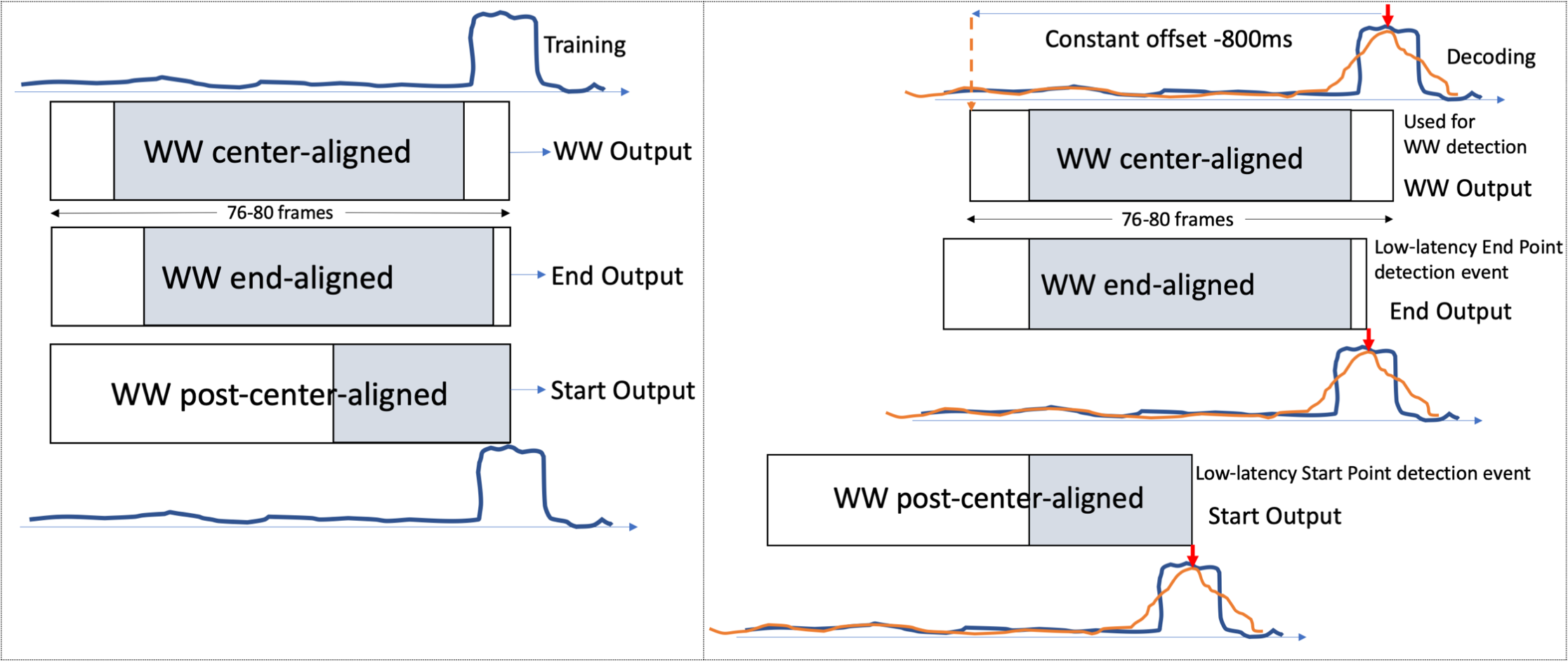}
\vspace{-6mm}
\caption{Endpointing in WL KWS using multi-aligned output model. (i) Training. (ii) Inference.}
\label{fig:start-end-align-training}
\vspace{-3mm}
\end{figure}

In this method, we train the WL KWS with additional outputs that detect different alignments of WW inside the input window, see Fig. \ref{fig:start-end-align-training} (i). That is, one output of the CNN may be detecting WW centrally positioned inside the input window, while another output may be detecting WW positioned to start at a given frame in the input window, yet another may be detecting WW positioned to end at a given frame of the input window. The time of the peaks of each output's posteriors then allows us to detect the WW center, start, and end time points, see Fig \ref{fig:start-end-align-training} (ii). 
\vspace{-1mm}

\begin{figure}[!htb]
\centering
\includegraphics[width=0.44\textwidth]{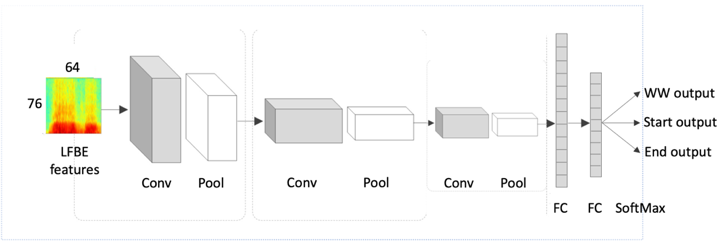}
\vspace{-3mm}
\caption{Architecture of the multi-aligned output WL KWS model.}
\label{fig:start-end-align-feats}
\vspace{-2mm}
\end{figure}

More specifically, we add two outputs to the main detector CNN output in the softmax layer, Fig. \ref{fig:start-end-align-feats}. The outputs are for detecting the start and end alignments of WW inside the input window. This is in addition to the main detector output, which is centrally aligned. We found such output to perform the best for the WW detection in experiments. It is also possible to use the information from all outputs together to generate WW detection events. However, we did not experiment with this option specifically.

\vspace{-3mm}

To train the model, we prepare WL training examples with WW differently aligned in the input window. An important point for start-aligned examples is that we align WW start with the middle of the input window, the 
``post-center-aligned" output 
in Fig. \ref{fig:start-end-align-training}.
Post-center alignment is introduced instead of a more straightforward WW alignment with the start of the input in order to reduce the latency of WW start detection. 
Because WL KWS has a long input window (e.g., 1 sec),  the start-aligned posterior may peak significantly after the WW end for WWs that are shorter than the WL input window.
We avoid this complication with post-center alignment. 
In this case, the WW start-aligned output becomes available before the other WW outputs. 
For end-aligned WW output, we prepare WL examples such that the WW end is aligned with the end of the input window up to a small margin, Fig. \ref{fig:start-end-align-training}.

We train the model by mixing differently aligned WL examples in training minibatch. Specifically, we used minibatch in proportions 25\%:12.5\%:12.5\%:50\% with respect to the center, start, end aligned, and negative WW examples, respectively, which we found to work the best in our experiments.
A larger weight for the centrally aligned WW in the minibatch is given to ensure better WW detection performance. 
We generate the minibatch in the described manner dynamically, during training. 
That is, WW examples are prepared first with the context of about 2 sec and central alignment of WW inside examples.
That allows selecting post-center and end-aligned WW examples later during training.
During training, the WW alignment is randomly chosen during the formation of the minibatch, according to the biased die above.
Random jitter is applied to the training examples by shifting the WW position slightly, to improve generalization. 
The examples are labeled in a one-hot manner according to their WW alignment and no-WW label.

\begin{table*}[!htbp]
\centering
\footnotesize
\begin{tabular}{|l|l|l|l|l|l|}
\hline
\multicolumn{1}{|c|}{\multirow{2}{*}{\textbf{Detector Name}}} &
  \multicolumn{2}{c|}{\textbf{All streams}} &
  \multicolumn{2}{c|}{\textbf{Long WW streams (length \textgreater 800 ms)}} &
  \multicolumn{1}{c|}{\multirow{2}{*}{\textbf{\begin{tabular}[c]{@{}c@{}}WW detection\\ (FRR at fixed FAR)\end{tabular}}}} \\ \cline{2-5}
\multicolumn{1}{|c|}{} &
  \multicolumn{1}{c|}{\textbf{Start Std Error}} &
  \multicolumn{1}{c|}{\textbf{End Std Error}} &
  \multicolumn{1}{c|}{\textbf{Start Std Error}} &
  \multicolumn{1}{c|}{\textbf{End Std Error}} &
  \multicolumn{1}{c|}{} \\ \hline
cnn\_const                                           &  46 ms & 107.2 ms & 55.7 ms & 116.7 ms & 9.3\%  \\ \hline
cnn\_regression\_multi\_task                                    &  17 ms & 100.1 ms & 18 ms & 75.6 ms & 10.23\%  \\ \hline
cnn\_regression\_thres\_crossing                                &  16.4 ms & 95.3 ms & \textbf{17.8 ms} & 73.3 ms & 9.3\%  \\ \hline
cnn\_regression\_local\_max                                     &  \textbf{16.2 ms} & 55 ms & 20.1 ms & 113.3 ms & 9.3\%  \\ \hline
cnn\_align                                                      &  22.4 ms & \textbf{40.9 ms} & 23.7 ms & \textbf{49 ms} & 9.3\%  \\ \hline
\end{tabular}
\caption{WW start end detection accuracy for different WW start-end detection models in this work.}{}
\label{tab:start-error}
\vspace{-8mm}
\end{table*}

\section{Experiments and results}
All models were trained for the keyword ``Alexa", which is the WW for the Echo family devices at Amazon.
We used 12M WW-positive and 5M WW-negative examples prepared from a corpus of annotated audio representing the far-field speech normally observed by Alexa devices 
in en-US locale.
For training either the WW start-end regression model or WW multi-aligned output model, the WW endpoints labels are necessary. Manually annotating WW endpoints in raw audio is extremely difficult and laborious. For that reason, we used WW endpoints generated by an AM+HMM KWS \cite{minhua1} as pseudo-ground truth for either training and evaluation.
Larger fragments of audio of 2 sec containing centered WW were prepared based on the AM+HMM KWS endpoints, as described. A set of negative examples comprising the AM+HMM KWS detections audio on data with negative annotations and central fragments of audio with negative annotations and without AM+HMM detections were added as the negative examples. 
WL CNN (Section 2) and Multi-aligned CNN were trained using cross-entropy loss and Adam optimizer in Tensorflow on that data for 2M steps using random initialization, mini-batch size of 4k, and learning rate of 0.001. The start-end regression model was trained for 50k steps using the frozen WL CNN model.

We tested the WW endpoints detection accuracy on held-out 33k streams. For that evaluation, we compare WW endpoints detection with pseudo-ground truth produced by the AM+HMM KWS, referred to as the ``online dataset" below.
Additionally, we evaluated WW start point accuracy versus human annotation on a smaller dataset. This dataset consists of 1100 ``Alexa" invocations in 5 different noise conditions, including household noises, external music, pink noise, and pink noise plus music. 
This dataset is referred to as the ``human-annotated dataset" below.
We also evaluated the WW detection accuracy with respect to possible degradation because of the addition of WW endpoints detection models.

These results are presented in Table \ref{tab:start-error}. The notation used in that table is as follows. 
The baseline \textit{cnn\_const} is the constant offset baseline described in the methods section. 
\textit{cnn\_align} is the multi-aligned outputs model.
\textit{cnn\_regression\_multi\_task} is the WW start-end regression model where the main CNN detector and the start-end model were trained simultaneously. 
\textit{thres\_crossing} and \textit{local\_max} are the same where the main CNN detector's weights were frozen for the training of the start-end regression model. 
\textit{thres\_crossing} and \textit{local\_max} are the versions of \textit{cnn\_regression} differing in the way the output of the start-end regression model is used during inference on streaming audio. 
Specifically, \textit{local\_max} uses the start-end outputs at the time where the raw WW posterior achieves local maximum, for calculating the WW endpoints. \textit{thres\_crossing} uses the first point of crossing by smooth posteriors of the detection threshold, for reduced detection latency. 
The metric used to quantify the accuracy of WW endpoints detection is the standard deviation of WW start or end errors (STD). This is calculated with respect to the AM+HMM endpoints in the ``online dataset" and human annotations in the ``human-annotated dataset". We present the WW endpoints accuracy split by WW length. This is because WW endpoints detection accuracy may be different for very long WWs. The metrics for the long WW category are calculated using ``Alexa" examples with WW duration exceeding 800 msec (1.3k streams out of 33k in ``online dataset"). WW duration of 800 msec is used here as the longest ten percentile for  ``Alexa". 

We now examine the results in Table \ref{tab:start-error}. For WW start, the multi-aligned outputs CNN makes a 60\% improvement vs. the \textit{cnn\_const} baseline. With the \textit{cnn\_regression\_local\_max} model, we obtain a 65\% improvement vs. \textit{cnn\_const}.  
In all models, the WW start accuracy with respect to the AM+HMM KWS on the ``online dataset" is about 20 msec. 

For WW end, multi-aligned output CNN makes a 65\% improvement vs. \textit{cnn\_const} baseline and close to 55\% improvement on long WWs. For \textit{cnn\_regression\_local\_max} model, we obtain a 49\% improvement vs. \textit{cnn\_const}. 
In absolute terms, multi-aligned output CNN provides WW end detection accuracy close to 40 msec and start-end regression model 55-110 msec, depending on the prescription for which time-point to use for the streaming start-end model outputs.

In summary, both proposed methods show superior accuracy for detecting the WW start in WL KWS. 
However, for WW end detection, the multi-aligned output CNN performs significantly better. 
One reason for that may be the uncertainty related to using the WW start-end regression model's outputs in streaming audio.
In streaming audio, the WW start-end regression model produces outputs continuously. 
Using different time-points for collecting the start-end outputs, which can be a smoothed posteriors' peak, raw posteriors' peak, first threshold crossing, leads to different WW endpoints prediction accuracy.
We find that using the start-end model's outputs when the raw WW posterior reaches its local maximum provided the best accuracy for WW start and end detections. 
However, this prescription leads to the worst performance on long WWs. 
See \textit{cnn\_regression\_local\_max} in  Table \ref{tab:start-error}.

We now discuss the impact of adding WW endpoint models on the accuracy of main WW detection, the last column in Table \ref{tab:start-error}. 
Metric used here is False Rejection Rate (FRR) at fixed False Acceptance Rate (FAR). 
In \textit{cnn\_regression\_thres\_cross} and \textit{cnn\_regression\_local\_max} detectors, since the weights of the detector CNN are frozen, the accuracy of WW detection is guaranteed to remain the same. 
However, if the WW start-end model is trained simultaneously with the main detector CNN, (\textit{cnn\_regression\_multi\_task}), we observe that WW detection accuracy can degrade by as much as relative 10\%. Multi-aligned output method, \textit{cnn\_align}, achieves similar WW detection accuracy as the standard CNN WL KWS.

Finally, we evaluated the WW start detection accuracy using a smaller human-annotated dataset.
Because ``start" annotation is laborious, This dataset is small. 
However, we can see in Table \ref{tbl:start-error-avs} that the accuracy of WW start detection for all our WL keyword spotters are, in fact, identical to that of the gold-standard AM+HMM KWS, on human-annotated data. 
This accuracy is about 50 msec STD. This accuracy is comparable, albeit worse, with the human annotators' accuracy, which we estimated to be about 30 msec. 

\begin{table}[!htb]
\scriptsize
\centering
\begin{tabular}{ | c | c |}
  \hline
  \textbf{Detector name} & \textbf{Start Std error}\\
  \hline
  AM+HMM & 52 ms\\
  \hline
  cnn\_align/regression & 52 ms\\
  \hline
  2nd\_human\_annotation & 31.2 ms\\
  \hline
\end{tabular}
\captionof{table}{WW start detection accuracy in WL KWS and 2-stage AM+HMM model using a smaller human-annotated dataset.}
\label{tbl:start-error-avs}
\vspace{-10mm}
\end{table}

\section{Conclusions and discussions}
We present two new methods for WW endpoints detection in the WL keyword spotter, namely the second WW start-end regression model and the multi-aligned output modeling method. 
In the case of the WW start point detection, the accuracy for ``Alexa" WW is found to be within 20 msec of the gold-standard AM+HMM keyword spotter and within 50 msec of human annotator, the latter being identical for the new methods and the industry gold-standard AM+HMM keyword spotter.
For the WW end point detection, the accuracy for ``Alexa" WW is found to be within 40 msec for multi-aligned output CNN and within 60 msec for the WW start-end regression model with respect to the gold-standard AM+HMM keyword spotter.


The multi-aligned output method provides overall superior accuracy for WW endpoints with a guaranteed KWS accuracy. It allows using a single CNN both for WW detection and WW start-end detection, thus, simplifying model training and inference.
On the other hand, the WW start-end regression model can guarantee that the WW detection accuracy remains constant with the WW endpoints model added, because it can be trained with the main detector KWS weights frozen.
In the multi-aligned output KWS, because the same feature representations are used for all WW alignments and WW detection, WW detection cannot be in principle decoupled from the endpoints detection, although we did not find that to create issues in practice.

Multi-aligned output models allow for early or low-latency detection of WW, i.e., by using the post-center aligned output. In particular, this output can produce a lower-accuracy WW detection by seeing a partial WW aligned post-center in the input window. 
Otherwise, the WL KWS incurs a constant latency of WW detection with respect, e.g., to the WW center point.
This feature is of great interest to the cascade KWS or VAD settings.
A unique advantage of multi-aligned outputs approach is also that it can be straightforwardly generalized to other applications, including ASR endpointing and WW detection using recurrent deep learning models such as LSTM. In this work, we considered a WW composed of a single keyword ``Alexa". It is straightforward to extend both presented methods to phrase WWs such as ``Ok Google".


\bibliographystyle{IEEEtran}


\bibliography{main-bibliography} 

\end{document}